\documentclass[referee]{raa}            

\usepackage{graphicx,times}             
\usepackage{natbib}
\usepackage{amssymb,amsmath}
\bibpunct{(}{)}{;}{a}{}{,}

\usepackage[pagebackref=true]{hyperref}
\hypersetup{colorlinks = true, linkcolor = green, anchorcolor = red, citecolor = blue, filecolor = red, pagecolor = red, urlcolor = red}

\begin{document}

   \title{ASAS J174406+2446.8 is identified as a marginal-contact binary with a possible cool third body
}

   \volnopage{Vol.0 (20xx) No.0, 000--000}      
   \setcounter{page}{1}          

   \author{Xiang-Dong Shi
      \inst{1,2,3,4}
   \and Sheng-Bang Qian
      \inst{1,2,3,4}
   \and Lin-Jia Li
      \inst{1,3,4}
   \and Wei-Wei Na
      \inst{5}
   \and Xiao Zhou
      \inst{1,3,4}
   }

   \institute{Yunnan Observatories, Chinese Academy of Sciences (CAS), P. O. Box 110, 650216 Kunming, China; {\it sxd@ynao.ac.cn}\\
        \and
             University of Chinese Academy of Sciences, Yuquan Road 19\#, Sijingshang Block, 100049 Beijing, China\\
        \and
             Key Laboratory of the Structure and Evolution of Celestial Objects, Chinese Academy of Sciences, P. O. Box 110, 650216 Kunming, China\\
        \and
             Center for Astronomical Mega-Science, Chinese Academy of Sciences, 20A Datun Road, Chaoyang District, Beijing, 100012, P. R. China\\
        \and
             Yuxi normal university, Fenghuang Road 134\#, 653100 Yuxi, China\\
\vs\no
   {\small Received~~20xx month day; accepted~~20xx~~month day}}

\abstract{  ASAS J174406+2446.8 was originally found as a $\delta$ Scuti-type pulsating star with the period P=0.189068 $days$ by ASAS survey. However, the LAMOST stellar parameters reveal that it is far beyond the red edge of pulsational instability strip on the $\log g-T$ diagram of $\delta$ Scuti pulsating stars. To understand the physical properties of the variable star, we observed it by the 1.0-m Cassegrain reflecting telescope at Yunnan Observatories. Multi-color light curves in B, V, R$_{c}$ and I$_{c}$ bands were obtained and are analyzed by using the W-D program. It is found that this variable star is a shallow-contact binary with an EB-type light curve and an orbital period of 0.3781\,days rather than a $\delta$ Scuti star. It is a W-subtype contact binary with a mass ratio of $1.135(\pm0.019)$ and a fill-out factor of $10.4(\pm5.6)\,\%$. The situation of ASAS J174406+2446.8 resembles those of other EB-type marginal-contact binaries such as UU Lyn, II Per and GW Tau. All of them are at a key evolutionary phase from a semi-detached configuration to a contact system predicted by the thermal relaxation oscillation theory. The linear ephemeris was corrected by using 303 new determined times of light minimum. It is detected that the O - C curve shows a sinusoidal variationthat could be explained by the light-travel-time effect via the the presence of a cool red dwarf. The present investigation reveals that some of the $\delta$ Scuti-type stars beyond the red edge of pulsating instability strip on the $\log g-T$ diagram are misclassified eclipsing binaries. To understand their structures and evolutionary states, more studies are required in the future.
}

   \authorrunning{X.-D. Shi, S.-B. Qian, L.-J. Li, W.-W. Na \& X. Zhou}            
   \titlerunning{ASAS J174406+2446.8 is detected as a marginal-contact binary with a possible cool third body }  

   \maketitle

%
%
\section{Introduction}           

A large number of variable stars including eclipsing binaries have been detected by several photometric sky surveys around the world, such as the Optical Gravitational Lensing Experiment Survey (OGLE; Udalski et al. 2015; Pietrukowicz et al. 2013), the Gaia mission (Gaia Collaboration et al. 2016, 2018), the All Sky Automated Survey (ASAS; Pojmanski 1997; Pojmanski, Pilecki \& Szczygiel 2005), the northern sky variability survey (NSVS; Wozniak et al. 2004), the Wide Angle Search for Planets (SuperWASP; Pollacco et al. 2006; Norton et al. 2011) and the Catalina Real-time Transient Survey (CRTS; Drake et al. 2012). Those detections have improved the development of the field of variable stars. However, the physical properties of many variable stars are unknown because of the lack of the spectroscopic data. During the first stage of spectroscopic survey of LAMOST (Wang et al. 1996; Cui et al. 2012), many spectroscopic data of variables were obtained by LAMOST including pulsating stars (Qian et al. 2018a, 2019a) and binary systems (Qian et al. 2017, 2018b). Moreover, more than two hundred and fifty thousand spectroscopic binary or variable star candidates were discovered by LAMOST (Qian et al. 2019b).

By using the LAMOST data, Qian et al. (2018a) constructed the $\log g-T$ diagram of $\delta$ Scuti pulsating stars that displayed in Fig. 1 and found a special group of $\delta$ Scuti stars(red dots in the figure, and blue dots refer of normal $\delta$ Scuti stars) are far beyond the red edge of pulsational instability strip on the diagram. These stars are distinguished from the normal $\delta$ Scuti stars, and may be a new-type pulsating star or may be misclassified. ASAS J174406+2446.8 (hereafter J1744) is one of those special pulsating stars and the position on the $\log g-T$ diagram is shown in Fig. 1. It is discovered as a $\delta$ Scuti-type pulsating star with a period of 0.189068\,days by the ASAS survey (Pojmanski 2002) and its related parameters are listed in Table 1. The ASAS light curve is shown in Fig. 2 that shows a large scatter indicating the classification may not reliable.The spectral atmospheric parameters of J1744 obtained by LAMOST are listed in Table 2. To study whether it is a new-type pulsating star or misclassified, we provide new CCD photometric light curves in BVR$_{c}$I$_{c}$ bands, then those light curves are carefully analyzed. At the same time, some new times of light minimum are obtained and the orbital periodic changes are investigated.

\begin{figure}\centering \vbox to4.0in{\rule{0pt}{5.0in}}
\includegraphics{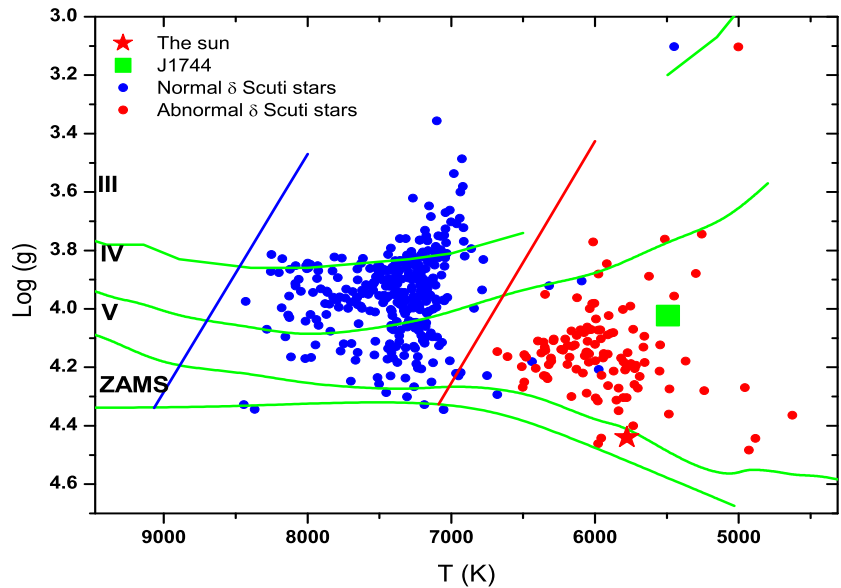}
\caption{ The relationships between Logg and T for $\delta$ Scuti-type pulsating stars observed by LAMOST (Qian et al. 2018a). Blue dots refer of normal $\delta$ Scuti stars, while red dots to unusual and cool $\delta$ Scuti stars.}
\label{}
\end{figure}

\begin{figure}\centering \vbox to4.0in{\rule{0pt}{5.0in}}
\includegraphics{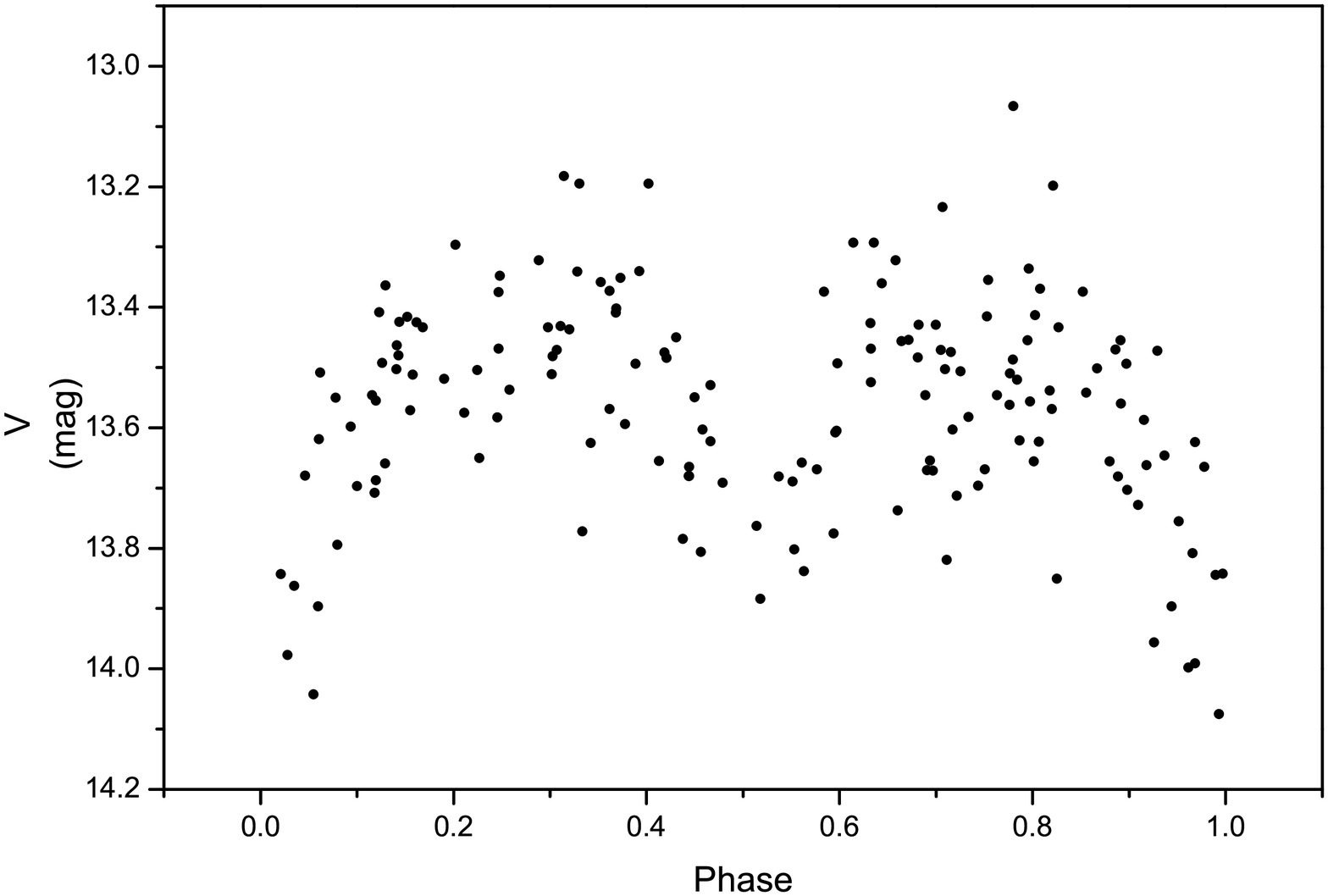}
\caption{ Light curve of J1744 observed by ASAS (The phase calculated by using the equation (1)).}
\label{}
\end{figure}

\begin{table}[]
\small
\begin{center} \caption{The information of J1744 from ASAS.}
\begin{tabular}{llllll}
 \hline\hline
      Parameters   & Value         \\
 \hline
      RA$_{2000}$  & 17 44 05.65   \\
      DEC$_{2000}$ & +24 46 47.1   \\
      V (mag)      & 12.85 (0.765) \\
      Period(days) & 0.189068      \\
\hline
\end{tabular}
\end{center}
\end{table}

\begin{table}[]
\small
\begin{center} \caption{The spectral information of J1744 released by LAMOST}
\begin{tabular}{llllllllllll}
 \hline\hline
       Parameters   & Value          \\
\hline
       Obsdate      & 2013-06-03     \\
       Subclass     & G7             \\
       Teff         & 5492.07(72.91) \\
       Log(g)       & 4.023(0.120)   \\
       $[Fe/H]$     & 0.186(0.070)   \\
       rv           & -21.00(12.00)  \\
\hline
\end{tabular}
\end{center}
\end{table}

\section{New CCD photometric observations}

In order to make a correct follow-up study, we observed the first CCD multi-color light curves for J1744. It was observed on May 15 and June 7, 2019 with the Andor DW-936N-BV CCD system on 1.0 m Cassegrain reflecting telescope(1m) at Yunnan Observatories (YNOs) in China. The telescopes have been equipped with the Johnson-Cousin-Bessel B, V, R$_{c}$ and I$_{c}$ filters. According to  weather conditions, period and band, the exposure time of J1744 is set to 50 seconds in B band, 40 seconds in V band, 30 seconds in R$_{c}$ band and 25 seconds in I$_{c}$ band. All the observed images were reduced by the aperture photometry package of the Image Reduction and Analysis Facility (IRAF) software. Differential magnitudes, which can represent the change of object star, were calculated from the magnitudes of object star minus the magnitudes of a nearby invariable comparison star. We tried to use twice the pulsation period given by ASAS to splice the phase of the light curve, in Fig. 3. \footnote{All of these photometric data are listed in the supplementary data} It was found that the magnitudes of the primary minima is obviously different to that in the secondary minima, the difference more than 0.1 mag, and the light variation out of eclipsing is continuous. That means J1744 is an EB-type eclipsing binary star, not a $\delta$ Scuti-type pulsating star.

\begin{figure}\centering \vbox to4.0in{\rule{0pt}{5.0in}}
\includegraphics{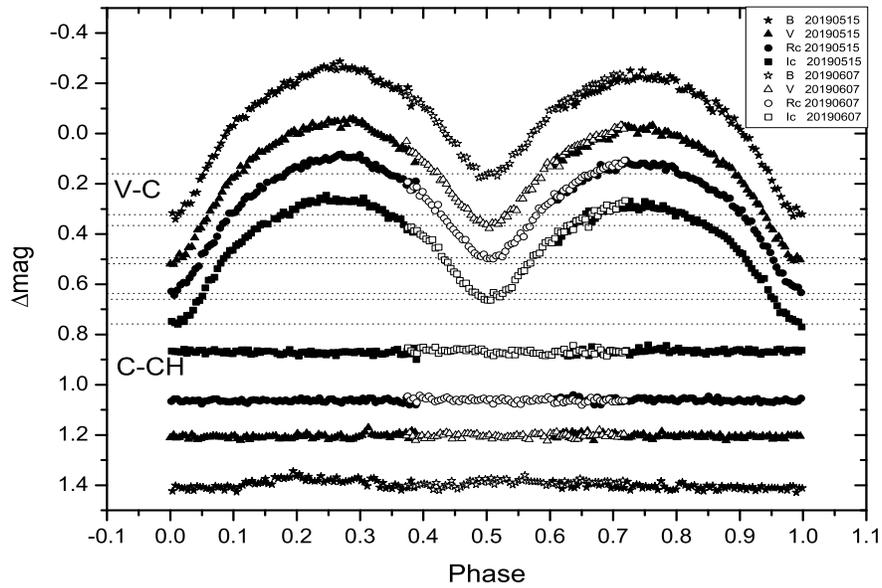}
\caption{ New CCD photometric observations in B, V, R$_{c}$ and I$_{c}$ of J1744 (The phase calculated by using the equation (1)).}
\label{}
\end{figure}

\section{The eclipse times and the orbital period analysis}

In order to get the eclipsing times of J1744, we collected photometric data from ASAS (Pojmanski 1997; Pojmanski 2002; Pojmanski, Pilecki \& Szczygiel 2005), NSVS (Wozniak et al. 2004), SuperWASP (Pollacco et al. 2006; Norton et al. 2011) and CRTS (Drake et al. 2012). By using the splicing phase and parabola fitting method, 303 light minimum are derived from the original data. \footnote{All of these eclipsing times data are listed in the supplementary data} Those eclipsing times include 158 primary times and 127 secondary times calculated using SuperWASP data, 3 primary times and 3 secondary times using ASAS data, 7 primary times and 3 secondary times using CRTS data and 1 primary times and 1 secondary times using our data. At here, a full SuperWASP raw data are kindly provided by Dr. M. E. Lohr, and the method of Obtaining eclipse time is borrowed from Lohr et al.  (2014).

By using the double period given by ASAS and a primary minimum, the linear ephemeris was derived as,
\begin{equation}
Min.I=2458619.26636+0^{d}.378136\times{E},
\end{equation}
the $(O - C)_1$ values of all available data were computed. In the upper panel of Fig. 4, the O - C values of all of light minimum show linear distributions indicating that the period needs to be revised. After linear fitting correction using all available data, a better linear ephemeris is determined as,
\begin{equation}
Min.I=2458619.27099(\pm0.00035)+0^{d}.378141051(\pm0.000000088)\times{E}.
\end{equation}
The $(O - C)_2$ diagram respect to the ephemeris is displayed in the middle panel of Fig. 4. As shown in the figure, the O - C curve display a cyclic change and the trend is more obvious if the more diffuse ASAS and CRTS data are removed, only reliable SuperWASP and 1m data are retained. The least-square method derived the following equation:
\begin{eqnarray}
(O - C)_2 &=& 0.0000015(\pm0.0003546)+0.0043(\pm0.0003)\nonumber\\
  &&\times{}sin\{\pi/5740.07273(\pm408.39194)\nonumber\\
  &&\times{[E-1384.08282(\pm936.86733)]}\}.\nonumber\\
\end{eqnarray}
In the equation, the sinusoidal term means a cyclic variation with a period of 11.89 years and an amplitude of 0.0043 days. The residuals fitting with equation (3) are shown in the lower panel of Fig. 4. After linear fitting to those residuals, the slope is $1.0\times{10^{-8}}$, which has less error than the orbital period in equation (2). So the equation (3) given a well fitting result.

\begin{figure}\centering \vbox to4.0in{\rule{0pt}{5.0in}}
\includegraphics{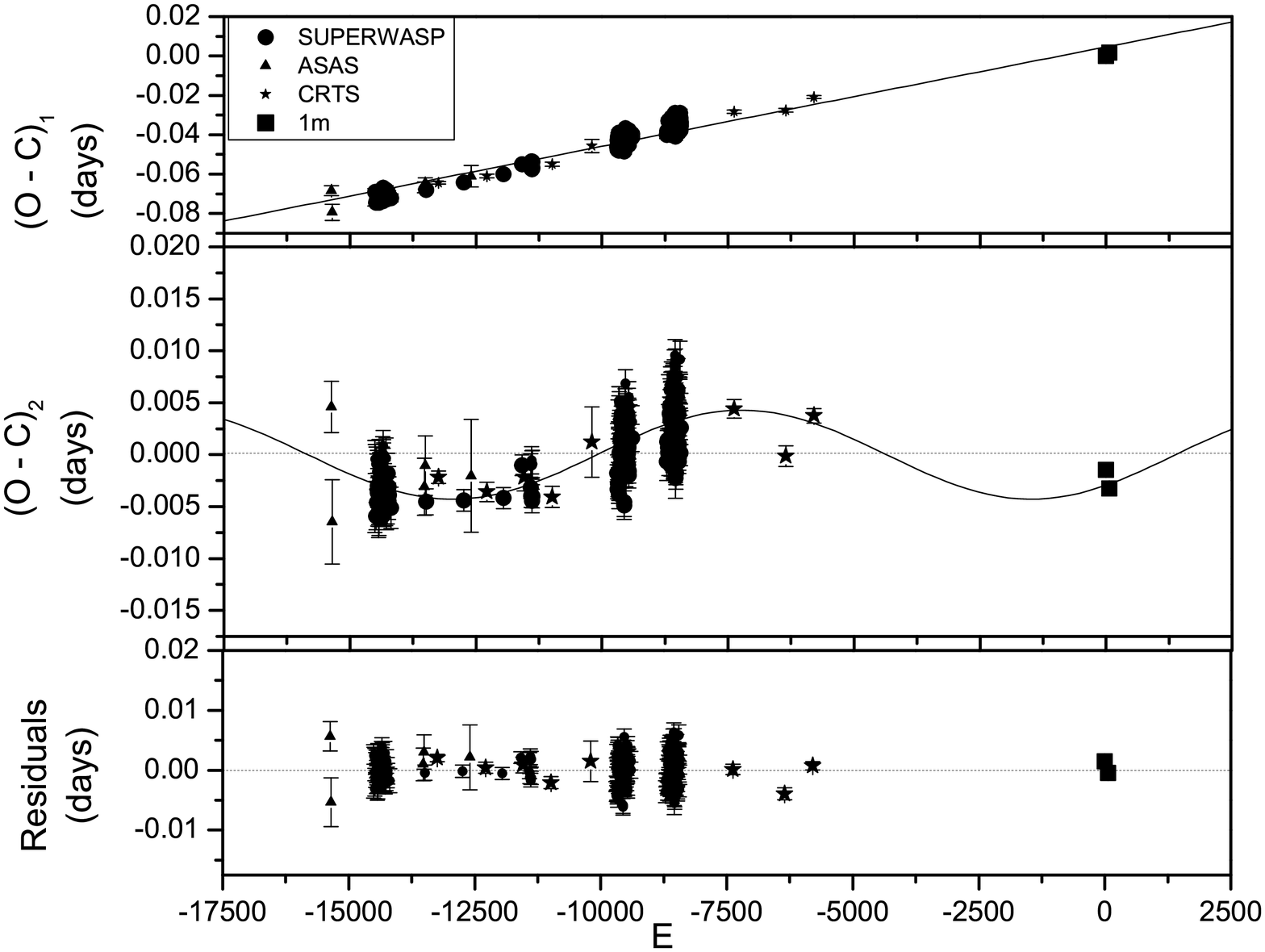}
\caption{ The O - C diagrams of the eclipsing binary J1744. The solid lines refer to the revised ephemeris.}
\label{}
\end{figure}

\section{Photometric solutions of J1744}

The first CCD multi-color light curves of J1744 was analyzed using the 2013 version of the W-D program (Wilson \& Devinney 1971; Wilson 1979, 1990, 2012 and Van Hamme \& Wilson 2007). According to spectral information, the temperature of star 1 was taken as $T_{1}=5492K$. It is a late-type binary and both components of the binary have convective envelope. Therefore, its bolometric albedo $A_{1}=A_{2}=0.5$ (Rucinski 1969) and its gravity-darkening coefficients $g_{1}=g_{2}=0.32$ (Lucy 1967) were assumed. To obtain the bolometric and bandpass limb-darkening coefficients, the parameter of limb darkening were taken according to the logarithmic law. To adjust the parameters (included $i$ (the orbital inclination); $T_{2}$ (the mean temperature of star 2); $q$ (the mass ratio of $M_2/M_1$); $L_{1}$(the monochromatic luminosity of star 1); and $\Omega_{1}$ (the dimensionless potential of star 1, and $\Omega_{1}=\Omega_{2}$ for overcontact configuration using mode 3) ), we will in principle obtain the solutions.

To obtain the highly critical parameter mass ratio q, a q-search method was performed for a series of mass ratios q from 0.2 to 6. The mode 2 (for detached binaries), mode 3 (for overcontact binaries), mode 4 (for semi-detached binaries with a lobe-filling star 1) and mode 5 (for semi-detached binaries with a lobe-filling star 2) were tried, but only at mode 3 it could converge. In Fig. 5, the relations between the sum of the squares of the residuals $\Sigma$ and the mass ratio q is displayed. And the minimum values of $\Sigma$ is at q = 3.2 from the numerical result and the parameters corresponding to this q value are listed in the second columns of Table 3. However, as shown in Fig. 5, the value of q from 1.2 to 4.7 the corresponding change of $\Sigma$ is very small. The observed data cannot be well fitted by the theoretical light curves, in Fig. 6, especially near to the 0.25 phase. The differences of magnitude between 0.25 phase and 0.75 phase is usually attributed to spot activity, such as, GR Tau (Qian 2002), V789 Her (Li et al. 2018), CN And (Van Hamme et al. 2001), V53 (Li et al. 2017), and LL Com (Hu et al. 2019) etc., where using the spot model the asymmetrical light curve could be successfully and reasonably fitted. In view of this situation that the 0.25 phase is brighter than the 0.75 phase, two spot models can be adopted: (i) a hot spot can be seen at phase near to 0.25; (ii) a cool spot can be seen at all the phase except near to 0.25. For the first model, the hot spot can be in either of the two sub-stars. For the second model, the cool spot must be observed at more phase than a hot spot in the first model, that means to achieve the same effect, a larger area of cool spot need to be used. After we test the two models, no matter which sub-star of J1744 the cool spot is on, only the coverage area more than $70\%$, the light curves could be fitted well. But using a hot spot, the coverage area is less than $30\%$, the light curves could also be fitted well. Both spot models are used for representation the non-uniformity of surface temperature and a large cool spot produces the same non-uniformity as a small hot spot on other remaining areas, So the second model is excluded. According to the location of the spot, there are two different situations for the first model: the hot spot on the sub-star 1 with a longitude of about $180^{\circ}$, or the hot spot on the sub-star 2 with a longitude of about $0^{\circ}$. With a hot spot on the sub-star 1, the q-search method was performed again, and the minimum values of $\Sigma$ is at q = 1.7, as shown in Fig. 7. Similarly, in Fig. 7 the minimum values of $\Sigma$ with a hot spot on the component 2 is at q = 1.2, and the values of $\Sigma$ is lower than that in the previous case. For the first case, a hot spot on the component 1, q = 1.7 as the initial value and as a free parameters, we fail to get a convergent solution. For the second case, a hot spot on the component 2, we obtained a good convergent solution near q = 1.2 for the same operation.

Therefore, the final solution is a hot spot on the component 2, and the parameters are listed in the third column of Table 3, corresponding theoretical and observed light curves are shown in Fig. 8. The geometrical structure of the 0.0, 0.25, 0.5 and 0.75 phase are shown in Fig. 9. The final solution shows that J1744 is a W-subtype contact binary star with a fill-out factor $f=10.4(\pm5.6)\,\%$ and a mass ratio $q=1.135(\pm0.019)$.

\begin{table}[]
\small
\begin{center} \caption{The Photometric Solutions of J1744 Using the W-D Code.}
\begin{tabular}{llllllllllll}
 \hline\hline
Parameters                                                          & Without spot              &  a hot spot on star 2    \\
\hline
Mode                                                                & Overcontact binary        &  Overcontact binary      \\
Gravity-darkening coefficients $(g_{1}=g_{2})^a$                    & 0.32(assumed)             &  0.32(assumed)           \\
Bolometric albedo $(A_{1}=A_{2})^a$                                 & 0.5(assumed)              &  0.5(assumed)            \\
Temperature of star 1 $(T_{1})^a$(K)                                & 5492(assumed)             &  5492(assumed)           \\
Orbital inclination i                                               & 71.44(18)                 &  69.847(85)              \\
Mass ratio q($M_2/M_1$)                                             & 3.2000000                 &  1.135(19)               \\
Temperature ratio $T_{2}/T_{1}$                                     & 0.9153(18)                &  0.9193(16)              \\
Luminosity ratio $L_{1}/(L_{1}+L_{2}) $ in band B                   & 0.3942(21)                &  0.6158(23)              \\
Luminosity ratio $L_{2}/(L_{1}+L_{2}) $ in band B                   & 0.6058(21)                &  0.3842(23)              \\
Luminosity ratio $L_{1}/(L_{1}+L_{2}) $ in band V                   & 0.3616(17)                &  0.5844(23)              \\
Luminosity ratio $L_{2}/(L_{1}+L_{2}) $ in band V                   & 0.6384(17)                &  0.4156(23)              \\
Luminosity ratio $L_{1}/(L_{1}+L_{2}) $ in band R$_{c}$             & 0.3432(14)                &  0.5657(22)              \\
Luminosity ratio $L_{2}/(L_{1}+L_{2}) $ in band R$_{c}$             & 0.6568(14)                &  0.4343(22)              \\
Luminosity ratio $L_{1}/(L_{1}+L_{2}) $ in band I$_{c}$             & 0.3302(12)                &  0.5522(22)              \\
Luminosity ratio $L_{2}/(L_{1}+L_{2}) $ in band I$_{c}$             & 0.6698(12)                &  0.4478(22)              \\
Modified dimensionless surface potential $\Omega_{1}=\Omega_{2}$    & 6.7943(89)                &  3.908(31)               \\
The degree of contact $(f)^b$                                       & 0.138(14)                 &  0.104(56)               \\
Radius of star 1 (relative to semimajor axis) in pole direction     & 0.26979(61)               &  0.3524(16)              \\
Radius of star 2 (relative to semimajor axis) in pole direction     & 0.45811(57)               &  0.3728(56)              \\
Radius of star 1 (relative to semimajor axis) in side direction     & 0.28175(73)               &  0.3707(18)              \\
Radius of star 2 (relative to semimajor axis) in side direction     & 0.49334(78)               &  0.3933(71)              \\
Radius of star 1 (relative to semimajor axis) in back direction     & 0.3189(12)                &  0.4056(19)              \\
Radius of star 2 (relative to semimajor axis) in back direction     & 0.52058(98)               &  0.427(11)               \\
Equal-volume radius of star 1 (relative to semimajor axis) $R_1$    & 0.29193(50)               &  0.3784(10)              \\
Equal-volume radius of star 2 (relative to semimajor axis) $R_2$    & 0.49189(45)               &  0.4006(46)              \\
Radius ratio $R_2/R_1$                                              & 1.6850(33)                &  1.059(12)               \\
Theoretical mean densities of star 1 $\rho_{1}(\rho_{\odot})^c$     & 0.8979(31)                &  0.8111(84)              \\
Theoretical mean densities of star 2 $\rho_{2}(\rho_{\odot})^c$     & 0.6006(35)                &  0.776(22)               \\
Latitude of spot $\theta$ (radian)                                  &   ...                     &  1.57079                 \\
Longitude of spot $\phi$ (radian)                                   &   ...                     &  0.19758                 \\
Angular radius of spot $r$ (radian)                                 &   ...                     &  0.46030                 \\
Dimensionless temperature factor of spot $T_{f}(T_{d}/T_{0})$       &   ...                     &  1.08731                 \\
$\Sigma{\omega(O - C)^2}$                                             & $4.71321\times{10^{-4}}$  & $2.91117\times{10^{-4}}$  \\
\hline
\end{tabular}
\end{center}
{$Notes.^a$  These were obtained by the method described in Section 4.\\
$^b$ $f=(\Omega_{star}-\Omega_{inner})/(\Omega_{outer}-\Omega_{inner})$, where $\Omega_{inner}, \Omega_{outer} and \Omega_{star}$ are
the modified dimensionless potential of inner Roche lobe, outer Roche lobe and star surface, respectively.\\
$^c$ The theoretical mean densities $M/(\frac{4} {3}R^3)$ in solar unit of 1410.040 842 Kg m$^{-3}$, which were derived from photometric solutions (Zhang et al. 2017).
Note. The two digital numbers in the parentheses are the errors on the last two bits of the data. }
\end{table}

\begin{figure}\centering \vbox to4.0in{\rule{0pt}{5.0in}}
\includegraphics{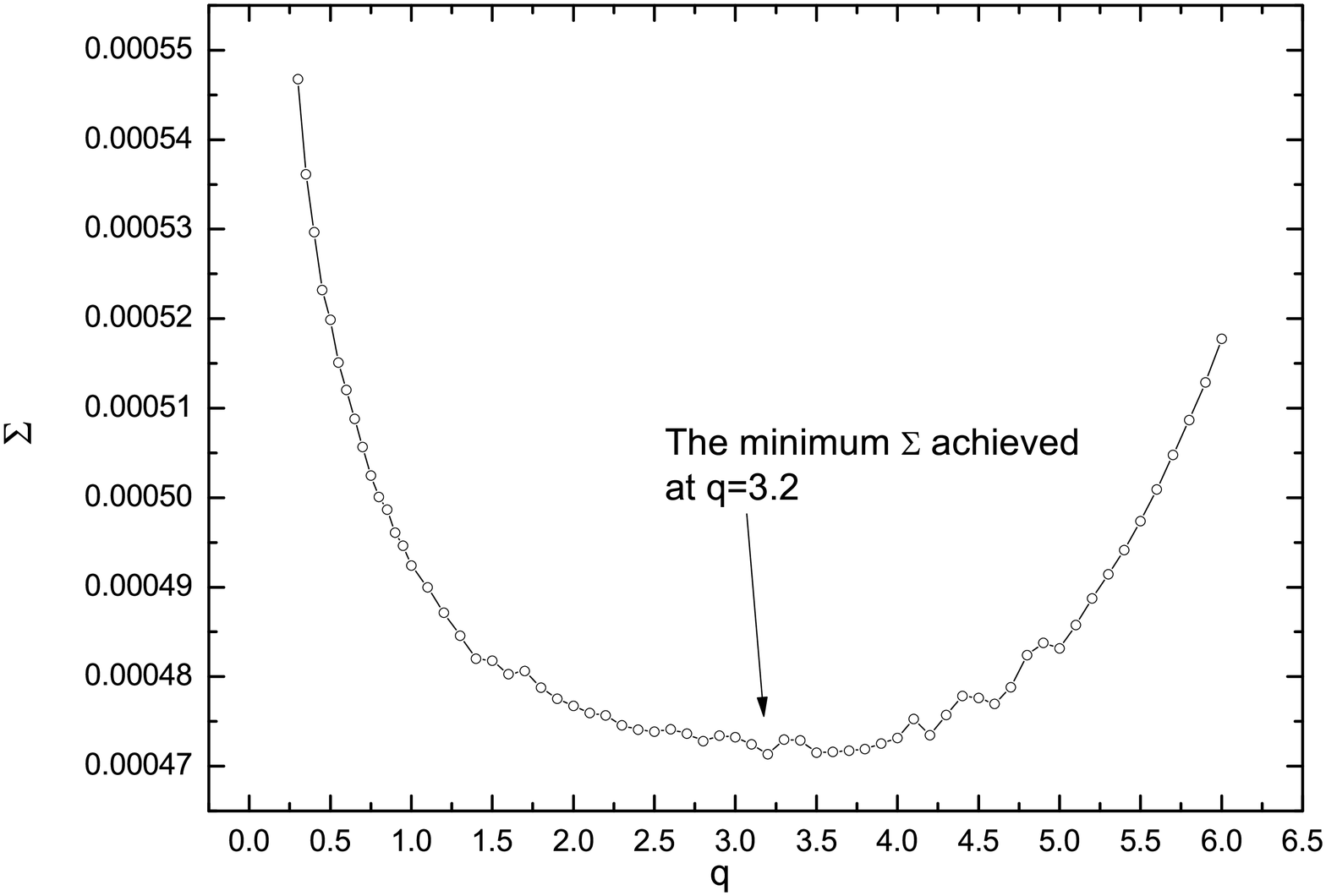} \caption{The relation between $\Sigma$ and q of J1744 without spot. }
\label{}
\end{figure}

\begin{figure}\centering \vbox to4.0in{\rule{0pt}{5.0in}}
\includegraphics{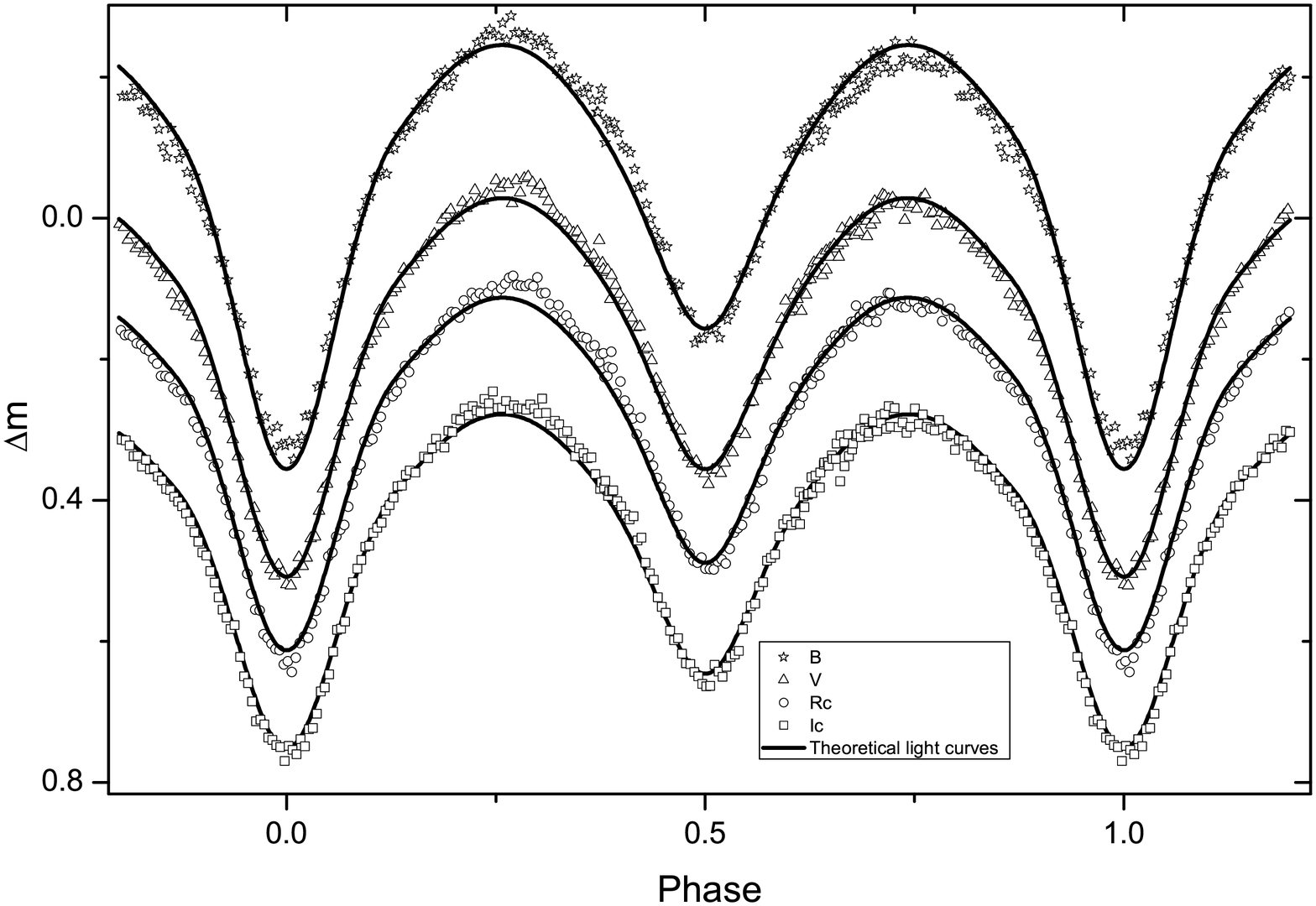} \caption{Theoretical (solid line) and observational light curves (black open stars, triangles, circles and squares) of J1744 without spot. }
\label{}
\end{figure}

\begin{figure}\centering \vbox to4.0in{\rule{0pt}{5.0in}}
\includegraphics{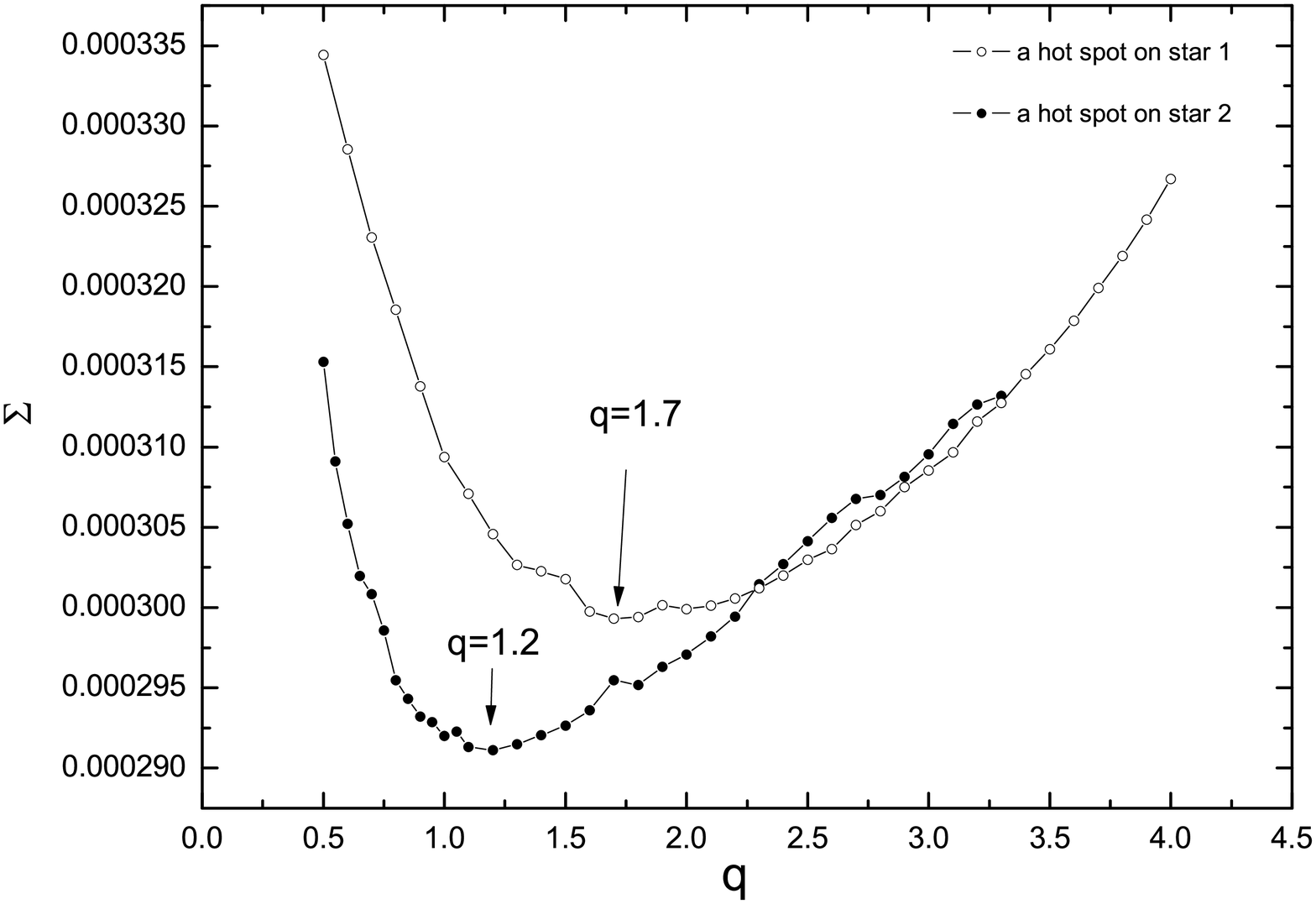} \caption{The relation between $\Sigma$ and q of J1744 with a hot spot on star 1 or star 2. }
\label{}
\end{figure}

\begin{figure}\centering \vbox to4.0in{\rule{0pt}{5.0in}}
\includegraphics{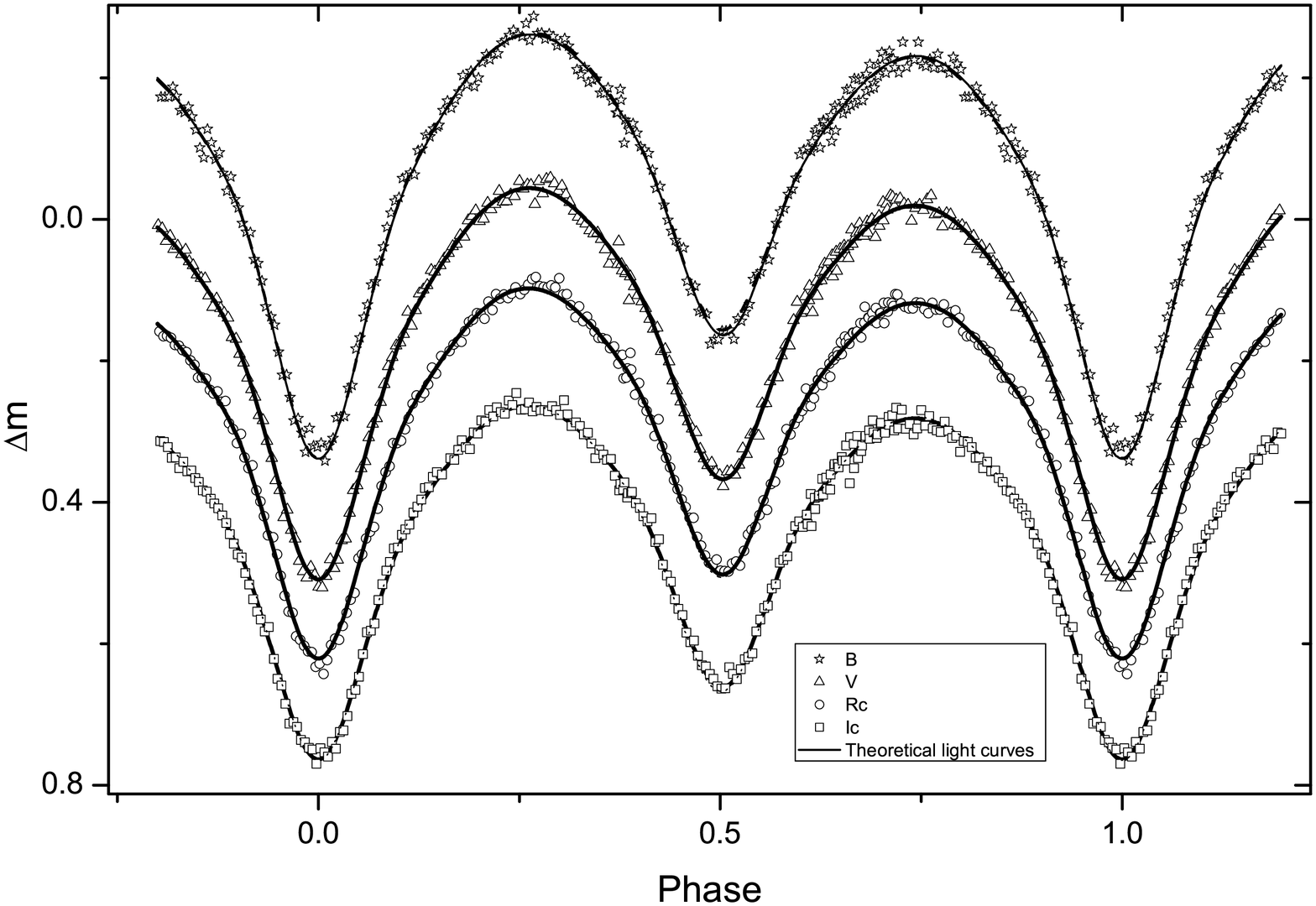} \caption{Theoretical (solid line) and observational light curves (black open stars, triangles, circles and squares) of J1744 with a hot spot on star 2. }
\label{}
\end{figure}

\begin{figure}\centering \vbox to4.0in{\rule{0pt}{5.0in}}
\includegraphics{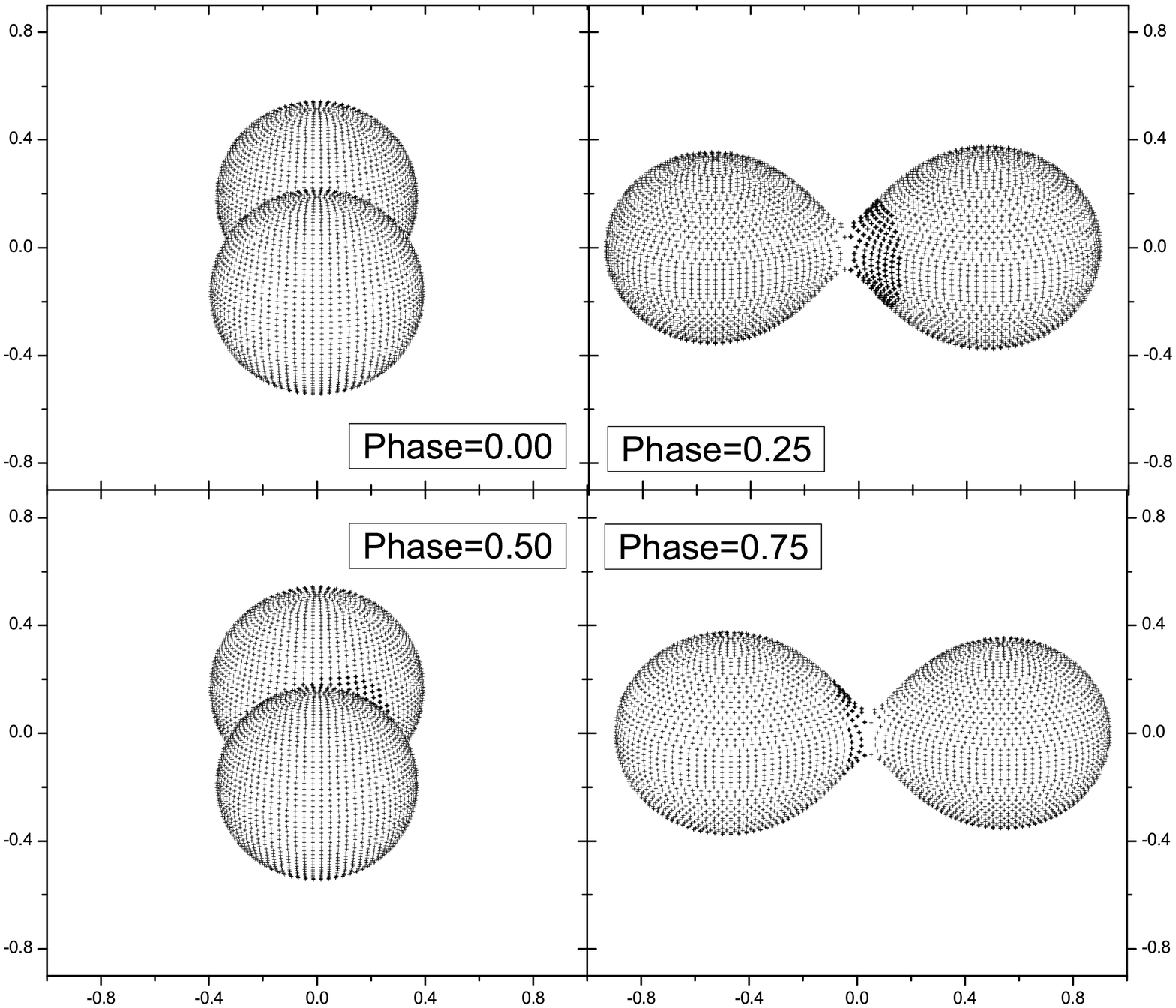} \caption{Geometrical structure of the overcontact binary system J1744 at phases 0.0, 0.25, 0.5, and 0.75 with a hot spot on star 2. }
\label{}
\end{figure}

\section{Discussions and Conclusions}

J1744 has been classified as a $\delta$ Scuti-type pulsating star with a period P=0.189068\,days. However, the LAMOST atmospheric parameters reveal that it is far beyond the red edge of pulsational instability strip of $\delta$ Scuti stars on the $\log g-T$ diagram (Qian et al. 2018a). This indicates that it is a special star or it was misclassified. To understand the observational properties of the star, we observed it by using the 1.0-m telescope at Yunnan Observatories. The more accurate multi-color light curves shown in Fig. 3 suggest that it is an EB-type eclipsing binary star with an orbital period of 0.3781\,days. The light varies continuously and the magnitude difference between the two minima is larger than 0.1\,mag.. By using the W-D program to analyze the light curves, it is detected that it is a low fill-out contact binary. The fill-out factor $f$ of the binary is $10.4(\pm5.6)\,\%$ and the mass ratio $q$ is $1.135(\pm0.019)$, where the surface temperature of the more massive component is slightly cooler than the less massive one.

According the theory of thermal relaxation oscillation (TRO) (Lucy 1976; Flannery 1976 and Robertson \& Eggleton 1977), contact binaries must undergo cycles around the state of marginal contact and they should be oscillating between the semi-detached and contact
configurations. However, the main problem of TRO remaining is the apparent non-existence of EB-type binaries with orbital periods shorter than 0.4\,days. (e.g., Rahunen 1981). The detection of J1744 as an EB-type binary with an orbital shorter than 0.4\,days (0.3781\,days) makes it an interesting system. It resembles UU Lyn (Zhu et al. 2007), II Per (Zhu et al. 2009), and GW Tau (Zhu \& Qian 2006) where both components are in the marginal contact. They may be at a key evolutionary phase from a semi-detached to a contact configurations and just at the beginning of contact phase as predicted by the TRO (Lucy 1976; Flannery 1976; Robertson \& Eggleton 1977; Lucy \& Wilson 1979).

Based on the spectral type of G7 given by LAMOST, the mass of the more massive component is estimated as $0.85M_{\odot}$ (Cox 2000), and combining $q=1.135$, so the mass of the less massive component is determined as $0.75M_{\odot}$. The light brightness at phase 0.25 is brighter than that of phase 0.75, which could be explained by a hot spot on the more massive but cooler component 2. From the photometric solutions, the surface temperature of component 1 is hotter than the component 2, and the mean density of component 1 is also higher than the component 2.

A total of 303 times of light minimum for the binary system are derived from all the photometric data collected. Firstly, by using these minimum times, we have corrected the linear ephemeris. Secondly, after the correction of the linear ephemeris, the O - C curve shows a cyclic period variation that can be well fitted by a sinusoidal curve. The sinusoidal variation means that the presence of a third body caused the light-travel time effect. We used the same method as Qian et al.(e.g., Qian et al. 2007, 2013) to determine
the parameters of the third body, and they are listed in Table 4. We try to search for the third light by using the W-D program, but its ratio to total light is always close to zero. All of this reveal that if a third body exists, the orbital inclination of the third body will not be very low, i.e., $i^{\prime}>30^{\circ}$, and that the third body is likely to be a cool red dwarf star, which cannot be observed because of its low luminosity contribution to the total system.

During the first stage of low-resolution spectroscopic survey, a total of 525 $\delta$ Scuti-type pulsating stars were observed by LAMOST (see Fig. 1). Qian et al. (2018a) constructed the $\log g-T$ diagram and found 131 abnormal $\delta$ Scuti stars that are far beyond the red edge of pulsating instability strip. As one of those abnormal $\delta$ Scuti stars, we find that J1744 is a misclassified short-period EB-type binary. It is a marginal-contact binary and just reaches the contact phase during the TRO cycles. To date, there are about 5575 $\delta$ Scuti-type pulsating stars included in the VSX (the international variable star index, e.g. Watson, Henden \& Price 2006). Our investigations indicate that some of them (or about 25\%) may be misclassified.
To understand the statistical properties of $\delta$ Scuti-type pulsating stars, they need to be investigated in details in the future.

\begin{table}[]
\small
\begin{center} \caption{The orbital parameters of the third companion in J1744.}
\begin{tabular}{llllllllllll}
 \hline\hline
Parameters &  Value & Units \\
\hline\hline
$A$ & $0.0043(\pm0.0003)$ & $days$\\
$P_{3}$ & $11.89(assumed)$ & $years$ \\
$a_{12}^{\prime}sini^{\prime}$ & $ 0.75(\pm0.05)$ & $A.U.$ \\
$f(m)$ & $2.93(\pm0.61)\times{10^{-3}}$ & $M_{\odot}$ \\
$m_{3}(i^{\prime}=90^{\circ})$ & $0.21(\pm0.02)$ & $M_{\odot}$ \\
$m_{3}(i^{\prime}=70^{\circ})$ & $0.23(\pm0.02)$ & $M_{\odot}$ \\
$m_{3}(i^{\prime}=50^{\circ})$ & $0.28(\pm0.02)$ & $M_{\odot}$ \\
$m_{3}(i^{\prime}=30^{\circ})$ & $0.46(\pm0.04)$ & $M_{\odot}$ \\
$a_{3}(i^{\prime}=90^{\circ})$ & $5.61(\pm0.58)$ & $A.U.$ \\
$a_{3}(i^{\prime}=70^{\circ})$ & $5.58(\pm0.58)$ & $A.U.$ \\
$a_{3}(i^{\prime}=50^{\circ})$ & $5.46(\pm0.57)$ & $A.U.$ \\
$a_{3}(i^{\prime}=30^{\circ})$ & $5.14(\pm0.56)$ & $A.U.$ \\
\hline
\end{tabular}
\end{center}
\end{table}

\begin{acknowledgements}
This work is supported by Chinese Natural Science Foundation (Nos. 11933008, 11703080 and 11803084), and the Yunnan Natural Science Foundation (No. 2018FB006). New CCD photometric observations of J1744 were obtained with the 1.0-m Cassegrain reflecting telescope (1m) at Yunnan Observatories (YNOs) in China. The photometric data obtained with ASAS, NSVS, CRTS and SuperWASP were used to calculate the eclipsing times. The spectral data of J1744 were provided by LAMOST. The authors would like to thank Dr. M. E. Lohr for kindly providing us the full SuperWASP original data.
\end{acknowledgements}

\label{lastpage}

\typeout{get arXiv to do 4 passes: Label(s) may have changed. Rerun}

\end{document}